\documentclass[aps,pra,twocolumn,showpacs,floatfix,nofootinbib,groupedaddress,superscriptaddress,citesort]{revtex4}
\usepackage{amsfonts}
\usepackage{mathbbold}
\usepackage{amstext}
\usepackage{amsmath}
\usepackage{amssymb}
\usepackage{upgreek}
\usepackage[dvips]{graphicx}
\def\qed{\leavevmode\unskip\penalty9999 \hbox{}\nobreak\hfill
     \quad\hbox{\leavevmode  \hbox to.77778em{%
               \hfil\vrule   \vbox to.675em%
               {\hrule width.6em\vfil\hrule}\vrule\hfil}}
     \par\vskip3pt}

\newtheorem{theorem}{Theorem}

\begin{document}
\title{Bound of Entanglement of Assistance and Monogamy Constraints}

\author{Zong-Guo Li}
\affiliation{Beijing National Laboratory for Condensed Matter
Physics, Institute of Physics, Chinese Academy of Sciences, Beijing
100080, China}
\author{Shao-Ming Fei}
 \affiliation{Department of Mathematics, Capital Normal University, Beijing 100037, China}
 \affiliation{Institut f{\"u}r Angewandte Mathematik, Universit{\"a}t Bonn, 53115, Germany}
\author{Sergio Albeverio}
\affiliation{Institut f{\"u}r Angewandte Mathematik, Universit{\"a}t
Bonn, 53115, Germany}
\author{W. M. Liu}
\affiliation{Beijing National Laboratory for Condensed Matter
Physics, Institute of Physics, Chinese Academy of Sciences, Beijing
100080, China}
\begin{abstract}
We investigate the entanglement of assistance which quantifies
capabilities of producing pure bipartite entangled states from a
pure tripartite state. The lower bound and upper bound of
entanglement of assistance are obtained. In the light of the upper
bound, monogamy constraints are proved for arbitrary n-qubit states.
\end{abstract}

\maketitle
\section{\bf Introduction}
In quantum information theory, entanglement is a vital resource for
some practical applications such as quantum cryptography, quantum
teleportation and quantum computation \cite{bennett,nielsen}. During
the last decade, this inspired a great deal of effort for detecting
and quantifying the entanglement
\cite{wootters,chen,mintert0,mintert1,gao,li1,ou,mintert2,li2}. On
the other hand, the creation and distribution of entanglement is
also of central interest in quantum information processing. More
specially the distribution of bipartite entanglement is a key
ingredient for performing certain quantum-information processing
tasks such as teleportation.

One of the methods for generating bipartite entanglement is the
entanglement of assistance that is defined in Refs. \cite{cohen,dp}.
It quantifies the entanglement which could be created by reducing a
multipartite entangled state to an entangled state with fewer
parties (e.g. bipartite) via measurements. Such producing of
entanglement, also called ``assisted entanglement", is a special
case of the \emph{localizable entanglement} \cite{localizable},
which is especially important for quantum communication, where
quantum repeaters are needed to establish bipartite entanglement
over a long length scale \cite{hj1}. For a pure $2\otimes2\otimes n$
state, the analytical formula of entanglement of assistance has been
derived by Laustsen \emph{et al.} \cite{laustsen}, whereas the
calculation of entanglement of assistance is not easy for a general
pure tripartite state \cite{gour}.

In this paper, we explore the entanglement of assistance for a
general pure tripartite state in terms of I-concurrence
\cite{rungta}. We obtain a lower bound of entanglement of
assistance, which is also the lower bound of a tripartite
entanglement measure, the entanglement of collaboration. This may
help to characterize the localizable entanglement. Furthermore, an
upper bound is also obtained. Deducing from the upper bound of
entanglement of assistance, we find a proper form of entanglement
monogamy inequality for arbitrary N-qubit states, which is analogous
to the monogamy constraints for concurrence proposed by Coffman
\emph{et al.} \cite{coffman} and proven by Osborne \emph{et al.}
 \cite{tobias} for the general case.

The paper is organized as follows: In Sec. II, we derive a lower
bound and upper bound of entanglement of assistance for pure
tripartite states. In Sec. III, monogamy constraints are proved in
terms of this upper bound. Finally in Sec. IV we conclude with a
discussion of our results.

\section{\bf Bound of entanglement of
Assistance} We consider a pure ($d_1\times d_2\times N$) tripartite
state shared by three parties referred to as Alice, Bob and Charlie,
who performs a measurement on his party to yield a known bipartite
entangled state shared by Alice and Bob. Charlie's aim is to
maximize the entanglement of the state between Alice and Bob. This
maximum average entanglement that he can create is called
entanglement of assistance, which was originally defined in terms of
entropy of entanglement \cite{dp,cohen}. In this paper, we define
entanglement of assistance in terms of the entanglement measure
I-concurrence:
\begin{eqnarray*}
E_a(|\psi\rangle_{ABC})\!\equiv\!
E_a(\rho_{AB})\!\equiv\!\textrm{max}\sum_ip_iC(|\phi_i\rangle_{AB}),
\end{eqnarray*}
which is maximized over all possible pure-state decompositions of
$\rho_{AB}=\textrm{Tr}_C[|\psi\rangle_{ABC}\langle\psi|]=\sum_ip_i|\phi_i\rangle_{AB}\langle\phi_i|$.
By applying the method in Ref. \cite{mintert0}, we can obtain the
lower bound of entanglement of assistance for pure tripartite
states.

For any given pure-state decomposition of $\rho_{AB}$,
$\rho_{AB}=\sum_ip_i|\phi_i\rangle_{AB}\langle\phi_i|$, we have
\begin{eqnarray}\label{lowbound eoa1}
E_a(|\psi\rangle_{ABC})\!&=&\!\textrm{max}\sum_ip_iC(|\phi_i\rangle_{AB})\nonumber\\
\!&=&\!\textrm{max}\sum_ip_i \sqrt{\sum_{mn}|\langle\phi_i|
S_{mn}|\phi_i^*\rangle|^2}\nonumber\\ \!&\geq&
\!\textrm{max}\sqrt{\sum_{mn}(\sum_ip_i |\langle\phi_i|S_{mn}
|\phi_i^*\rangle|)^2},
\end{eqnarray}
where $S_{mn}=L_m\otimes L_n$, $L_m, m=1,..., d_1(d_1-1)/2$, $L_n,
n=1, ..., d_2(d_2-1)/2$ are the generators of group $SO(d_1)$ and
$SO(d_2)$ respectively. The inequality holds according to the
Minkowski inequality $[ \sum\limits_{i=1}(
\sum\limits_{k}x_{i}^{k})^{p}] ^{1/p} \leq \sum_{k}[
\sum\limits_{i=1}( x_{i}^{k}) ^{p}] ^{1/p},\text{ }p>1$. Consider
the eigenvalue decomposition of $\rho_{AB}$, $\rho_{AB}=\Psi M\Psi
^{\dagger }$, where $M$ is a diagonal matrix whose diagonal elements
are the eigenvalues of $\rho$, and $\Psi$ is a unitary matrix whose
columns are the eigenvectors of $\rho $. Taking into account the
relation $\Phi W^{1/2}=\Psi M^{1/2}U$, where $U$ is a right-unitary
matrix, we can rewrite inequality (\ref{lowbound eoa1}) as
\begin{eqnarray*}
\!\!E_a(\rho_{AB})\!&\geq&\!\textrm{max}\sqrt{\sum_{mn}(\sum_i
|\Phi^T
W^{\frac{1}{2}} S_{mn}W^{\frac{1}{2}}\Phi|_{ii})^2}\nonumber\\
\!&=&\!\textrm{max}\sqrt{\sum_{mn}(\sum_i |U^T M^{\frac{1}{2}}\Psi^T
S_{mn}\Psi M^{\frac{1}{2}}U|_{ii})^2}.
\end{eqnarray*}

In terms of the Cauchy-Schwarz inequality $(\sum_i
x_i^2)^{\frac{1}{2}}(\sum_iy_i^2)^{\frac{1}{2}}\geq\sum_i x_iy_i$,
the inequality
\begin{eqnarray}
\label{lowbound eoa2}
E_a(\rho_{AB})\!\geq\!\textrm{max}\sum_i\left|U^T\left(\sum_{mn}z_{mn}A_{mn}\right)U\right|_{ii}
\end{eqnarray}
is implied for any $z_{mn}=y_{mn}exp(i\theta_{mn})$ with
$y_{mn}\geq0$ and $\sum_{mn}y_{mn}^2=1$, where
$A_{mn}=M^{\frac{1}{2}}\Psi^T S_{mn}\Psi M^{\frac{1}{2}}$. Since
$\sum_{mn}z_{mn}A^{mn}$ is a symmetric matrix, we can always find a
unitary matrix $U$ such that
$\sum_i|U^T(\sum_{mn}z_{mn}A_{mn})U|_{ii}=\|\sum_{mn}z_{mn}A^{mn}\|$
as shown in Ref. \cite{horn}, where $\|\cdot\|$ stands for the trace
norm defined by $\|G\|=\textrm{Tr}(GG^\dag)^{1/2}$. For an arbitrary
unitary matrix $V$, we have
\begin{eqnarray*}
&&\sum_i|V^T(\sum_{mn}z_{mn}A_{mn})V|_{ii}\\&\!\!=\!\!&\sum_i|V^T
(U^{-1})^TU^T(\sum_{mn}z_{mn}A_{mn})UU^{-1}V|_{ii}\\
&\!\!=\!\!&\sum_i|V^T
(U^{-1})^TDiag(\lambda_1,\lambda_2\cdots)U^{-1}V|_{ii}\\
&\!\!\leq\!\!&\sum_{ij}|(U^{-1}V)_{ij}|^2\lambda_i\\
&\!\!=\!\!&\sum_i\lambda_i,
\end{eqnarray*}
where $\lambda_{i}(z)$s, dependent
on the choice of the $y$ and $\theta$, are the singular values of
the matrix $\mathcal {T}=\sum_{mn}z_{mn}A^{mn}$, i.e., the square
roots of the eigenvalues of the positive Hermitian matrix $\mathcal
{T}\mathcal {T}^\dag$. Therefore the maximum of Eq. (\ref{lowbound
eoa2}) is given by $\underset{z\in
\mathbf{C}}{max}\left(\sum_i\lambda_{i}(z)\right)=\underset{z\in
\mathbf{C}}{max}\|\sum_{mn}z_{mn}A^{mn}\|$. Hence, we arrive at the
lower bound of entanglement of assistance for a pure tripartite
state as following:
\begin{eqnarray}
\label{bound of eoa} E_a(\rho_{AB})\!\geq\!\underset{z\in
\mathbf{C}}{\textrm{max}}\|\sum_{mn}z_{mn}A^{mn}\|.
\end{eqnarray}

Furthermore the entanglement of collaboration \cite{gour2,gour3}
quantifies the maximum amount of entanglement that can be generated
between two parties from a tripartite state with collaborations
composed of local operations and classical communication among the
three parties. It has been shown by Gour \emph{et. al.} \cite{gour2}
that, for tripartite states, the entanglement of collaboration is
greater than or equal to entanglement of assistance in terms of a
given entanglement measure. Therefore our lower bound is also the
one for entanglement of collaboration, which can be tightened by
numerical optimization. Our bound may help to characterize
localizable entanglement. For a pure $2\times2\times N$ state, this
lower bound is consistent with the result of Ref. \cite{laustsen}.

We can also obtain the upper bound of entanglement of assistance.
From the definition of entanglement of assistance, we have
\begin{eqnarray*}
[E_a(\rho_{AB})]^2&=&[\textrm{max}\sum_ip_iC(|\phi_i\rangle_{AB})]^2\nonumber\\
&\!\leq\!&\textrm{max}\sum_i[\sqrt{p_i}C(|\phi_i\rangle_{AB})]^2\sum_i(\sqrt{p_i})^2\nonumber\\
&\!=\!&\textrm{max}\sum_i2p_i[1-\textrm{Tr}(\rho_i^A)^2]\nonumber\\
&\!\leq\!&2(1-\textrm{Tr}\rho_A^2),
\end{eqnarray*}
where $\rho_i^A=\textrm{Tr}_B|\phi_i\rangle_{AB}\langle\phi_i|$. The
first inequality holds according to the Cauchy-Schwarz inequality
\cite{tj}; the last one, which has also been proved in Ref.
\cite{vicente}, holds due to the convex property of
$\textrm{Tr}\rho_A^2$.

Define the upper bound as the tangle of assistance
$\tau_a(\rho_{AB})\equiv\textrm{max}\sum_ip_i[C(|\phi_i\rangle_{AB})]^2$.
Similar to the entanglement of assistance that satisfies the
monogamy constraints for n-qubit pure state \cite{gour1,monogamy},
we show below that the tangle of assistance also exhibits monogamy
constraints for arbitrary n-qubit states.

\section{\bf Monogamy inequality}
Consider a pure tripartite state $|\Psi\rangle_{ABC}$. The tangle of
assistance is defined by
\begin{eqnarray*}
\tau_a(|\Psi\rangle_{ABC})&=&\underset{\{p_x,|\psi_x\rangle\}}{\textrm{max}}\sum_xp_x[C(|\psi_x\rangle)]^2\\
&=&\underset{\{p_x,|\psi_x\rangle\}}{\textrm{max}}\sum_xp_xS_2[\textrm{Tr}_B(|\psi_x\rangle\langle\psi_x|)],
\end{eqnarray*}
where the linear entropy $S_2[\rho]=2[1-\textrm{Tr}(\rho)^2]$, and
the maximum runs over all pure-state decompositions
$\{p_x,|\psi_x\rangle\}$ of
$\rho_{AB}=\textrm{Tr}_C(|\Psi\rangle_{ABC}\langle\Psi|)=\sum_xp_x|\psi_x\rangle\langle\psi_x|$.
In the case of pure state $\rho_{AB}$, the tangle of assistance is
the square of concurrence of this state.
\begin{theorem}
For an arbitrary n-qubit state, the tangle of assistance satisfies,
\begin{eqnarray}
\label{monogamy}
&&\tau_a(\rho_{A_1A_2})+\tau_a(\rho_{A_1A_3})+\cdots+\tau_a(\rho_{A_{1}A_n})\nonumber\\&\geq&\tau_a(\rho_{A_1(A_2A_3\cdots
A_n)}),
\end{eqnarray}
where $\tau_a(\rho_{A_1(A_2A_3\cdots A_n)})$ denotes the tangle of
assistance in the bipartite partition $A_1|A_2A_3\cdots A_n$.
\end{theorem}


 {\em Proof:} First of all, we prove the following inequality
 \begin{equation}
 \label{monogamy1}
\tau_a(\rho_{AB})+\tau_a(\rho_{AC})\geq\tau_a(\rho_{A(BC)}),
 \end{equation}
for arbitrary tripartite states $\rho_{ABC}$ in
$2\times2\times2^{n-2}$ system.

We first prove Eq. (\ref{monogamy1}) for pure states. In this case,
due to the local-unitary invariance of $\tau_a(\rho_{AC})$, we can
rotate the basis of subsystem $C$ into the local Schmidt basis
$|V_k\rangle$, $k=1,\cdots,4$, given by the eigenvectors of
$\rho_C=Tr_{AB}(\rho_{ABC})$. In this way we can regard the
$2^{n-2}$-dimensional qudit $C$ as an effective four-dimensional
qudit. Therefore, we simply need to prove Eq. (\ref{monogamy1}) for
a $2\times2\times4$ pure state $ABC$.

For pure states of a tripartite system $ABC$ of two qubits $A$ and
$B$ and a four-level system $C$, we have
\begin{eqnarray*}
&&\tau_a(\rho_{A(BC)})-\tau_a(\rho_{AC})\\&=&S_2(\rho_A)
-\underset{\{p_j,|\phi_j\rangle\}}{\textrm{max}}\sum_jp_jS_2[\textrm{Tr}_C(|\phi_j\rangle\langle\phi_j|)],
\end{eqnarray*} where $\sum_jp_j|\phi_j\rangle\langle\phi_j|=\rho_{AC}$. It can be
shown that any pure-state decomposition of $\rho_{AC}$ can be
realized by positive-operator-valued measures (POVMs) $\{M_x\}$
performed by Bob, the rank of which is 1 (for more details see
\cite{gour,lp}). Therefore, we get the the following expression
\begin{eqnarray}
\label{def1}
\tau_a(\rho_{AC})=\underset{\{M_x\}}{\textrm{max}}\sum_xp_xS_2(\rho_x),
\end{eqnarray}
where the maximum runs over all rank-1 POVMs on Bob's system,
$p_x=\textrm{Tr}(I_A\otimes M_x\rho_{AB})$ is the probability of
outcome $x$, and $\rho_x=\textrm{Tr}_B(I_A\otimes M_x\rho_{AB})/p_x$
is the posterior state in Alice's subsystem. For convenience, we
take the definition
\begin{eqnarray*}
I(\rho_{AB}):=S_2(\rho_A)-\underset{\{M_x\}}{\textrm{max}}\sum_xp_xS_2(\rho_x).
\end{eqnarray*}
By comparing $I(\rho_{AB})$ with Eq. (\ref{monogamy1}) for pure
tripartite states, we see that it is sufficient to prove the
inequality
\begin{eqnarray*}
I(\rho_{AB})\leq\tau_a(\rho_{AB}),
\end{eqnarray*}
for all two-qubit states $\rho_{AB}$.

We first derive a computable formula for $I(\rho_{AB})$. Any
bipartite quantum  state $\rho_{AB}$ may be written as
\begin{eqnarray}
\label{formaula1} \rho_{AB}=\Lambda\otimes
I_B(|V_{B'B}\rangle\langle V_{B'B}|),
\end{eqnarray}
where $V_{B'B}$ is the symmetric two-qubit purification of the
reduced density operator $\rho_B$ on an auxiliary qubit system $B'$
and $\Lambda$ is a qubit channel from $B'$ to $A$. Deducing from Eq.
(\ref{def1}) we have
\begin{eqnarray*}
\rho_x&\!\!=\!\!&\textrm{Tr}_B(I_A\otimes M_x\rho_{AB})/p_x\\
&\!\!=\!\!&\textrm{Tr}_B[(I_A\otimes M_x)(\Lambda\otimes
I_B)|V_{B'B}\rangle\langle
V_{B'B}|)]/p_x\\
&\!\!=\!\!&\Lambda[\textrm{Tr}_B(I_A\otimes
M_x|V_{B'B}\rangle\langle V_{B'B}|)]/p_x.
\end{eqnarray*}
Since the
rank of $M_x$ is 1, $\textrm{Tr}_B(I_A\otimes
M_x|V_{B'B}\rangle\langle V_{B'B}|)]$ is a pure state. Moreover, all
pure-state decompositons of
$\rho_B'=\textrm{Tr}_B(|V_{B'B}\rangle\langle V_{B'B}|)=\rho_B$ can
be realized by the rank-1 POVM measurements $\{M_x\}$ operating on
subsystem $B$ of $|V_{B'B}\rangle\langle V_{B'B}|$. Hence
$I(\rho_{AB})$ satisfies
\begin{equation}
\label{def2}
I(\rho_{AB})=S_2[\Lambda(\rho_B)]-\underset{\{p_x,|\psi_x\rangle\}}{\textrm{max}}\sum_xp_xS_2[\Lambda(|\psi_x\rangle)],
\end{equation}
where the maximum runs over all pure-state decompositions
$\{p_x,|\psi_x\rangle\}$ of $\rho_B$ such that
$\sum_xp_x|\psi_x\rangle\langle\psi_x|=\rho_B$.

The action of a qubit channel $\Lambda$ on a single-qubit state
$\rho=(I+\mathbf{r}\cdot\boldsymbol{\upsigma})/2$, where
$\boldsymbol{\upsigma}$ is the vector of Pauli operators, may be
written as
$\Lambda(\rho)=[I+(\mathbf{L}\mathbf{r}+\mathbf{l})\cdot\boldsymbol{\upsigma}]/2$,
where $\mathbf{L}$ is a $3\times3$ real matrix  and $\mathbf{l}$ is
a three-dimensional vector. In this Pauli basis, the possible
pure-state decompositions of $\rho_B$ are represented by all
possible sets of probabilities $\{p_j\}$ and unit vectors
$\{\mathbf{r}_j\}$ such that $\sum_jp_j\mathbf{r}_j=\mathbf{r}_B$,
where $(I+\mathbf{r}_B\cdot\boldsymbol{\upsigma})/2=\rho_B$. In
terms of the Block representation of one-qubit states, the linear
entropy $S_2$ is given by
$S_2[(I+\mathbf{r}\cdot\boldsymbol{\upsigma})/2]=1-|\mathbf{r}|^2$.
In this way we get the following equation
$S_2[\Lambda(I+\mathbf{r}\cdot\boldsymbol{\upsigma})/2]
=1-(\mathbf{L}\mathbf{r}+\mathbf{l})^T(\mathbf{L}\mathbf{r}+\mathbf{l})$.

Substituting $\mathbf{r}_j=\mathbf{r}_B+\mathbf{x}_j$, one can
easily check that Eq. (\ref{def2}) reduces to the following one
whose value is determined by $\{p_j,\mathbf{x}_j\}$ subject to the
conditions $\sum_jp_j\mathbf{x}_j=0$ and
$|\mathbf{r}_B+\mathbf{x}_j|=1$,
\begin{eqnarray}
\label{def3} &\!&I(\rho_{AB})\nonumber\\
&\!=\!&S_2[\Lambda(\rho_B)]-\underset{\{p_j,\mathbf{x}_j\}}{\textrm{max}}
\sum_jp_jS_2[\Lambda(\frac{I+(\mathbf{r}_B+\mathbf{x}_j)\cdot\boldsymbol{\upsigma}}{2})]\nonumber\\
&\!=\!&1-(\mathbf{L}\mathbf{r}_B+\mathbf{l})^T(\mathbf{L}\mathbf{r}_B+\mathbf{l})\nonumber\\
&\!-\!&\underset{\{p_j,\mathbf{x}_j\}}{\textrm{max}}\sum_j
p_j\Big\{1-[\mathbf{L}(\mathbf{r}_B+\mathbf{x}_j)+\mathbf{l}]^T[\mathbf{L}(\mathbf{r}_B+\mathbf{x}_j)+\mathbf{l}]\Big\}\nonumber\\
&=&\underset{\{p_j,\mathbf{x}_j\}}{\textrm{min}}\sum_j
p_j(\mathbf{x}^T_j\mathbf{L}^T\mathbf{L}\mathbf{x}_j).
\end{eqnarray}
Without loss of generality, we assume that $\mathbf{L}^T\mathbf{L}$
is diagonal with diagonal elements
$\lambda_x\leq\lambda_y\leq\lambda_z$. The constrains
$|\mathbf{r}_B+\mathbf{x}_j|=1$ lead to the identities
$(\mathbf{x}^x_j)^2=1-|\mathbf{r}_B|^2-2\mathbf{r}_B^T\mathbf{x}_j-(\mathbf{x}^y_j)^2-(\mathbf{x}^z_j)^2$.
Substituting this into Eq. (\ref{def3}), we get
$I(\rho_{AB})=\lambda_x(1-|\mathbf{r}_B|^2)+\underset{\{p_j,\mathbf{x}_j\}}{\textrm{min}}
\sum_jp_j[(\lambda_y-\lambda_x)(\mathbf{x}^y_j)^2+(\lambda_z-\lambda_x)(\mathbf{x}^z_j)^2]$.
This expression is obviously minimized by choosing
$\mathbf{x}^z_j=\mathbf{x}^y_j=0$ for all $j$. Then from the
condition $|\mathbf{r}_B+\mathbf{x}_j|=1$, $\mathbf{x}^x_j$ have two
solutions. The ensemble of two states corresponding to such two
solutions can reach the minimum $\lambda_x(1-|\mathbf{r}_B|^2)$.

As $S_2(\rho_B)=(1-|\mathbf{r}_B|^2)$, we obtain the following
computable expression: $I(\rho_{AB})=\lambda_{min} S_2(\rho_B)$.
Note that a local filtering operation of the form
$\rho'_{AB}=\frac{(I\otimes B)\rho_{AB}(I\otimes
B^\dag)}{\textrm{Tr}[(I\otimes B^\dag B)\rho_{AB}]}$ leaves
$\mathbf{L}$ invariant and transforms
$S_2(\rho_{B'})=\frac{\textrm{det}(B)^2}{\textrm{Tr}[(I\otimes
B^\dag B)\rho_{AB}]^2}S_2(\rho_{B})$  \cite{frank}.

If the local filtering
 operator $B$ is invertible, we can get the conclusion that there does not
exist a pure-state decomposition $\{q_j,|\psi_j\rangle\}$ of
$\rho'_{AB}$ such that
$\tau_a(\rho'_{AB})>\frac{\textrm{det}(B)^2}{\textrm{Tr}[(I\otimes
B^\dag B)\rho_{AB}]}\tau_a(\rho_{AB})$ by the contradiction. For the
case that the operator $B$ is not invertible, such pure-state
decomposition also doesn't exist. Furthermore, there exists exactly
an optimal pure-state decomposition $\{p_i,|\phi_i\rangle\}$ of the
state $\rho_{AB}$ for $\tau_a(\rho_{AB})$ such that
$\sum_ip_iC[\frac{(I\otimes B)(|\phi_i\rangle\langle\phi_i|I\otimes
B^\dag)}{\textrm{Tr}[(I\otimes B^\dag
B)\rho_{AB}]}]^2=\frac{\textrm{det}(B)^2}{\textrm{Tr}[(I\otimes
B^\dag B)\rho_{AB}]^2}\tau_a(\rho_{AB})$. Therefore, the tangle of
assistance
$\tau_a(\rho'_{AB})=\frac{\textrm{det}(B)^2}{\textrm{Tr}[(I\otimes
B^\dag B)\rho_{AB}]^2}\tau_a(\rho_{AB})$. Since
$I(\rho'_{AB})=\frac{\textrm{det}(B)^2}{\textrm{Tr}[(I\otimes B^\dag
B)\rho_{AB}]^2}\lambda_{min}S_2(\rho_{B})$, it transforms exactly in
the same way as the tangle of assistance $\tau_a(\rho'_{AB})$ does.
As there always exists a filtering operation for which
$\rho_B'\propto I$, we can assume, without loss of generality, that
$S_2(\rho_B)=1$.

So let us consider $\rho_{AB}$ with
$\rho_B=\textrm{Tr}_A(\rho_{AB})=\frac{1}{2}I$. In terms of Pauli
operators, we can rewrite the pure state as follows:
\begin{eqnarray*}
&&\frac{(I\otimes B)|V_{B'B}\rangle\langle V_{B'B}|(I\otimes
B^\dag)}{\textrm{Tr}[(I\otimes B^\dag B)|V_{B'B}\rangle\langle
V_{B'B}|]}\\
&\!\!=\!\!&\frac{1}{4}[I+\sum_im_iI\otimes\sigma_i+\sum_in_i\sigma_i\otimes
I+\sum_{ij}O_{ij}\sigma_i\otimes\sigma_j],
\end{eqnarray*}
 where $\sigma_1$,
$\sigma_2$ and $\sigma_3$ are $\sigma_x$, $\sigma_y$ and $\sigma_z$
respectively. Then we get the conclusion from its purity and unity
reduced density, that $m_i=n_i=0$ for all i and the $3\times3$ real
matrix $O$ is orthogonal. Thus we have
$\rho_{AB}=\frac{1}{4}\Lambda\otimes
I_B[I+\sum_{ij}O_{ij}\sigma_i\otimes\sigma_j]
=\frac{1}{4}[I+\sum_il_i\sigma_i\otimes
I+\sum_{ij}(LO)_{ij}\sigma_i\otimes\sigma_j]$. As unitary operator
$U_1$ satisfies the equation
$U_1\sigma_iU_1^\dag=\sum_jP_{ij}\sigma_j$, where $P$ is a real
orthogonal $3\times3$ matrix, we can always find local unitary
operators, in terms of the theorem of singular value decomposition,
so that $U_1\otimes U_2\rho_{AB}U_1^\dag\otimes
U_2^\dag=\frac{1}{4}[I+\sum_i(lP)_i\sigma_i\otimes
I+\sum_{ij}(QLOP)_{ij}\sigma_i\otimes\sigma_j]=\frac{1}{4}[I+\sum_il'_i\sigma_i\otimes
I+\sum_{i}(L')_{ii}\sigma_i\otimes\sigma_i]$, where $Q$ and $P$ are
real orthogonal matrix and $L'$ is a diagonal matrix with its
diagonal elements the singular values of $L$.
Because of the local-unitary invariance of $\tau_a(\rho_{AB})$ and
$I(\rho_{AB})$, without loss of generality, we assume that
$\rho_{AB}=\frac{1}{4}[I+\sum_it_i\sigma_i\otimes
I+\sum_{i}(R)_{ii}\sigma_i\otimes\sigma_i]$, where $R$ is a diagonal
matrix with its diagonal elements the singular values of $L$. Due to
the positivity of
\begin{eqnarray*}
\rho_{AB}=
  \!\!\!\frac{1}{4}\left(\!\!\!
   \begin{array}{cccc}
   1+R_3+t_3\!\!\!&\!\!\!    0     \!\!\!&\!\!\!t_1-it_2   \!\!\!&\!\!\!R_1-R_2   \\
   0        \!\!\!&\!\!\!1-R_3+t_3 \!\!\!&\!\!\!R_1+R_2    \!\!\!&\!\!\!t_1-it_2  \\
   t_1+it_2 \!\!\!&\!\!\!R_1+R_2   \!\!\!&\!\!\!1-R_3-t_3  \!\!\!&\!\!\!0         \\
   R_1-R_2  \!\!\!&\!\!\!t_1+it_2  \!\!\!&\!\!\!0          \!\!\!&\!\!\!1+R_3-t_3 \\
   \end{array}
     \!\!\right)\!\!,
\end{eqnarray*}
the inequality $1-t_1^2-t_2^2-t_3^2\geq R_3^2$ must hold. Therefore
we obtain
\begin{eqnarray*}
\!\!\!&&\!\!\!\tau_a(\rho_{AB})\geq[C_a(\rho_{AB})]^2\\
\!\!\!&\geq&\!\!\!\textrm{Tr}[\sigma_y\otimes\sigma_y\rho^*_{AB}\sigma_y\otimes\sigma_y\rho_{AB}]\\
\!\!\!&=&\!\!\!\frac{1}{16}\left[4+4(R_1^2+R_2^2+R_3^2)-4(t_1^2+t_2^2+t_3^2)\right]\\
\!\!\!&\geq&\!\!\!\frac{1}{4}[R_1^2+R_2^2+2R_3^2]\\
\!\!\!&\geq&\!\!\!\lambda_{min}(\mathbf{L}^T\mathbf{L}).
\end{eqnarray*}
This inequalities imply that $I(\rho_{AB})\leq\tau_a(\rho_{AB})$ for
all two-qubit states $\rho_{AB}$, which then proves Eq.
(\ref{monogamy1}) for pure states.

Now we extend Eq. (\ref{monogamy1}) to mixed state case. Consider
the maximizing pure-state decomposition $\{p_x,|\psi_x\rangle\}$ for
$\tau_a(\rho_{A(BC)})$. By applying the inequality Eq.
(\ref{monogamy1}) and taking into account the concavity of $\tau_a$,
we have
\begin{eqnarray*}
\tau_a(\rho_{A(BC)})&=&\sum_xp_x\tau_a(\rho^x_{A(BC)})\\
&\leq&\sum_xp_x[\tau_a(\rho^x_{AB})+\tau_a(\rho^x_{AC})]\\
&\leq&\tau_a(\rho_{AB})+\tau_a(\rho_{AC}),
\end{eqnarray*}
where $\rho^x_{A(BC)}=|\psi_x\rangle\langle\psi_x|$.

Let $C=C_1C_2$ be a $2\times2^{n-3}$ system and apply Eq.
(\ref{monogamy1}), then we get
\begin{eqnarray*}
\tau_a(\rho_{A(BC)})\!\!\!&&\!\!\!\leq\tau_a(\rho_{AB})+\tau_a(\rho_{AC})\\
\!\!\!&&\!\!\!\leq\tau_a(\rho_{AB})+\tau_a(\rho_{AC_1})+\tau_a(\rho_{AC_2}).
\end{eqnarray*}
 Successively applying Eq.
(\ref{monogamy1}) to partitions of $C$, we obtain the inequality Eq.
(\ref{monogamy}) by induction. $\blacksquare$

In fact, Eq. (\ref{monogamy}) turns out to be an equality for
product states under partition $A|BC_1\cdots C_n$. For the
generalized GHZ states, Eq. (\ref{monogamy}) is a strictly
inequality.

\section{\bf Discussion}
In summary, as an important quantity in quantum computation, the
entanglement of assistance has been investigated in terms of
I-concurrence for pure tripartite states.  We have obtained a lower
bound of entanglement of assistance, which is also the lower bound
of the tripartite entanglement measure, the entanglement of
collaboration. In stead of great difficulty involved in computing
the entanglement of collaboration, the lower bound Eq. (\ref{bound
of eoa}) can be calculated in a numerical optimization to make a
good estimation of entanglement of collaboration. Moreover, an upper
bound is also obtained. In the light of the upper bound of
entanglement of assistance, we find a proper form of entanglement
monogamy inequality for arbitrary N-qubit states.

This work was supported by NSFC under grants Nos. 60525417,
10740420252, 10874235, 10875081, 10675086, the NKBRSFC under grants
Nos. 2006CB921400, 2009CB930704, KZ200810028013 and
NKBRPC(2004CB318000).


\begin{thebibliography}{99}
\bibitem{bennett} C. H. Bennett and D. P. DiVincenzo, Nature (London) \textbf{404}, 247
(2000).
\bibitem{nielsen} M. A. Nielsen and I. L. Chuang, \emph{Quantum Computation and Quantum Information}
(Cambridge University Press, Cambridge, 2000).
\bibitem{wootters} W. K. Wootters, Phys. Rev. Lett. \textbf{80}, 2245 (1998).
\bibitem{mintert0} F. Mintert, M. Ku\'{s}, and A. Buchleitner, Phys. Rev. Lett. \textbf{92}, 167902 (2004).
\bibitem{mintert1} F. Mintert, M. Ku\'{s}, and A. Buchleitner, Phys. Rev. Lett. \textbf{95}, 260502 (2005).
\bibitem{chen} K. Chen, S. Albeverio, and S. M. Fei, Phys. Rev. Lett. \textbf{95}, 040504 (2005).
\bibitem{gao} X. H. Gao, S. M. Fei, and K. Wu, Phys. Rev. A \textbf{74}, 050303(R) (2006).
\bibitem{li1} Z. G. li, F. S. Fei, Z. X. Wang and K. Wu, Phys. Rev. A \textbf{75}, 012311 (2007)
\bibitem{ou} Y.  C. Ou, H. Fan, and S. M. Fei, Phys. Rev. A \textbf{7}8, 012311 (2008).
\bibitem{mintert2} L. Aolita, A. Buchleitner, and  F. Mintert, Phys. Rev. A \textbf{78}, 022308 (2008).
\bibitem{li2} Z. G. li, F. S. Fei, Z. D. Wang and W. M. Liu, Phys. Rev. A \textbf{79}, 024303 (2009)
\bibitem{dp} D. P. DiVincenzo, C. A. Fuchs, H. Mabuchi, J. A. Smolin, A. Thapliyal, and A. Uhlmann,
\emph{The Entanglement of assistance}, Lecture Notes in Computer
Science Vol. 1509 (Springer-Verlag, Berlin, 1999), pp. 247-257
\bibitem{cohen} O. Cohen, Phys. Rev. Lett. \textbf{80}, 2493 (1998).
\bibitem{localizable} F. Verstraete, M. Popp, and J. I. Cirac, Phys. Rev. Lett. \textbf{92}, 027901
(2004); M. Popp, F. Verstraete, M. A. Martin-Delgado, and J. I.
Cirac, Phys. Rev. A \textbf{71}, 042306 (2005).
\bibitem{hj1} H. J. Briegel, W. D\"{u}r, J. I. Cirac, and P. Zoller, Phys. Rev. Lett. \textbf{81}, 5932 (1998).
\bibitem{laustsen} T. laustsen, F.Berstraete, and S. J. van Enk, Quantum Inf. Comput. \textbf{3}, 64 (2003).
\bibitem{gour} G. Gour, Phys. Rev. A \textbf{72}, 042318 (2005).
\bibitem{rungta} P. Rungta, V. Bu\v{z}ek, C. M. Caves, M. Hillery, and G. J. Milburn, Phys. Rev. A \textbf{64},
042315 (2001).
\bibitem{coffman} V. Coffman, J. Kundu, and W. K. Wootters, Phys. Rev. A \textbf{61},
052306 (2000).
\bibitem{tobias} T. J. Osborne and F. Verstraete, Phys. Rev. Lett. \textbf{96}, 220503 (2006).
\bibitem{horn} R. A. Horn and C. R. Johnson, \emph{Matrix Analysis} (Cambridge University Press, New York, 1985), p. 205.
\bibitem{gour2} G. Gour  and R. W. Spekkens, Phys. Rev. A \textbf{73}, 062331 (2006).
\bibitem{gour3} G. Gour, Phys. Rev. A \textbf{74}, 052307 (2006).
\bibitem{tj} T. J. Osborne, Phys. Rev. A \textbf{72}, 022309 (2005).
\bibitem{vicente} J. I. de Vicente, J. Phys. A: Math. Theor. \textbf{41}, 065309 (2008).
\bibitem{gour1} G. Gour, D. A. Meyer, and B. C. Sanders, Phys. Rev. A \textbf{72}, 042329 (2005).
\bibitem{monogamy} G. Gour, S. Bandyopadhyay, and B. C. Sanders, J. Math. Phys.
48, \textbf{012108} (2007).
\bibitem{lp} L. P. Hughston, R. Jozsa, and W. K. Wootters, Phys. Lett. A
\textbf{183}, 14 (1993).
\bibitem{frank} F. Verstraete, J. Dehaene, and B. DeMoor, Phys. Rev. A \textbf{64}, 010101(R)
(2001).

\end{thebibliography}
\end{document}